\documentclass{article}
\usepackage{amsfonts}
\usepackage{amssymb}
\usepackage{amsmath}
\usepackage{amsthm}
\usepackage[english]{babel}
\begin{document}

\title{\bf On the recently proposed\\ Mimetic Dark Matter}

\author{{\bf Alexey Golovnev}\\
{\small {\it Saint-Petersburg State University, high energy physics department,}}\\
{\small \it Ulyanovskaya ul., d. 1; 198504 Saint-Petersburg, Petrodvoretz; Russia}\\
{\small agolovnev@yandex.ru, \quad golovnev@hep.phys.spbu.ru}}
\date{}

\maketitle

\begin{abstract}

Recently, an interesting gravitational model was proposed in order to mimic the effect of Dark Matter. Chamseddine and Mukhanov in the arXiv preprint 1308.5410 have separated the conformal mode of a physical metric in the form of a squared gradient of an auxiliary scalar field. Notably, the variational principle has given a more general equation of motion than that of purely Einsteinian relativity theory, with a possibility of reproducing an effective Dark Matter. In this short paper, we explain the nature of this phenomenon in terms of the class of functions on which the variation takes place. Then we give a more transparent equivalent formulation of the model without an auxiliary metric. Finally, we speculate a bit about possible extensions.

\end{abstract}

\vspace{1cm}

\section{Introduction}

One of the biggest puzzles in modern cosmology is the nature of Dark Matter which persistently evades any kind of unequivocal detection outside the realm of gravitational interactions at galactic and cosmological scales. Not surprising, many attempts were made to model its observed effects by a suitable type of modified gravitational interaction, ranging from MOND \cite{MOND} to Ho{\v r}ava gravity \cite{Horava}.

Yet another, very interesting, model has been proposed in Ref. \cite{mimetic}. The idea is to represent the physical metric $g_{\mu\nu}$ in the Einstein-Hilbert variational principle as
\begin{equation}
\label{definition}
g_{\mu\nu}={\tilde g}_{\mu\nu}{\tilde g}^{\alpha\beta}(\partial_{\alpha}\phi)(\partial_{\beta}\phi)
\end{equation}
with an auxiliary metric ${\tilde g}_{\mu\nu}$ and a scalar field $\phi$ of unusual dimension $-1$. The conformal mode of ${\tilde g}$ is stripped off any physical significance, and its role is seemingly relegated to the scalar player. 

However, it follows from (\ref{definition}) that, apart from the equation of motion
\begin{equation}
\label{eom}
\bigtriangledown_{\mu}\left((G-T)\partial^{\mu}\phi\right)=0,
\end{equation}
the scalar field identically satisfies the constraint
\begin{equation}
\label{fixednorm}
g^{\alpha\beta}(\partial_{\alpha}\phi)(\partial_{\beta}\phi)=1.
\end{equation}
And after substituting it into the Einstein equations,
\begin{equation}
\label{Einstein}
G_{\mu\nu}-T_{\mu\nu}-(G-T)(\partial_{\mu}\phi)(\partial_{\nu}\phi)=0
\end{equation}
with $G_{\mu\nu}$ and $T_{\mu\nu}$ being the Einstein tensor and the matter stress tensor respectively, and the trace quantites denoted by letters without indices, we see that the trace part of equations is satisfied identically.

Therefore, we have an extra freedom in the system. And effectively there is an extra contribution to the stress tensor of the form
\begin{equation}
\label{effective}
{\tilde T}_{\mu\nu}=(G-T)(\partial_{\mu}\phi)(\partial_{\nu}\phi)
\end{equation}
which amounts to introducing a pressureless dust with potential flow, or the {\it Mimetic Dark Matter}, with initial energy density being a mere integration constant \cite{mimetic}.

In Section 2 we explain this outcome from the viewpoint of variational calculus. In Section 3 we present a simpler formulation of the same model given by the action (\ref{final}). And, finally, we present some speculations and conclude in Section 4.

\section{The structure of variational principle}

It might be surprising that only by rearranging the parts of the metric, without introducing any new ingredients into the action, we end up with a new model at hand. Let us however look at this more attentively. Without loss of generality, assume that ${\tilde g}$ is unimodular. Therefore, the determinant of $g$ must be given by (the fourth power of) the factor of
$$\Omega(x)\equiv{\tilde g}^{\alpha\beta}(\partial_{\alpha}\phi)(\partial_{\beta}\phi).$$
To reproduce the standard Einstein equations, we need to vary the action 
$$-\int d^4x R(g({\tilde g},\phi))$$
with respect to the metric $g$ including the $\Omega$ factor with the only restriction that the variation be vanishing at the boundary of the variation domain, most importantly at the initial and the final instants of time. This is not ensured by our definition of $\Omega$.

For the sake of simplicity, assume a spatially homogeneous variation $\phi(t)=t+\delta\phi(t)$ around a spatially homogeneous field $\phi=t$ in a space-time of the form $ds^2=dt^2-a^2(t)d{\overrightarrow x}^2$. We have $\dot\phi=\sqrt{\Omega}$. Given that $\delta\phi=0$ at $t=t_{in}$ and $t=t_{fin}$, we obtain $\int\limits_{t_{in}}^{t_{fin}}dt\sqrt{\Omega}=t_{fin}-t_{in}$ for any admissible variation. It shows that we perform the variation in a restricted class of functions. Such variation provides less conditions for the stationarity of the action, and therefore there is more freedom in dynamics.

This is nothing more but a general problem of making derivative substitutions into the action. Suppose we have an action $S=\int{\dot x}^2 dt$, and substitute $x\equiv {\dot y}$ into it. After that, the equations of motion are of higher order, even for $x(t)$, and therefore one needs more Cauchy data. The reason is the same as above. We vary in the class of functions which is defined not only by vanishing of $\delta x$ at the boundary but also by an extra $\int\limits_{t_{in}}^{t_{fin}}\delta x(t) dt=0$ condition.

\section{Equivalent formulation}

Now we want to find an equivalent formulation of the mimetic DM model. As a first step, we introduce a set of Lagrange multipliers $\lambda^{\mu\nu}$ and write an action
$$S=-\int\left(R(g)+\lambda^{\mu\nu}\left(g_{\mu\nu}-{\tilde g}_{\mu\nu}{\tilde g}^{\alpha\beta}(\partial_{\alpha}\phi)(\partial_{\beta}\phi)\right)\right)\sqrt{-g}d^4x.$$
The matter Lagrangian is totally omitted. However, if needed, its effect can be restored by simply subtracting $T_{\mu\nu}$ form $G_{\mu\nu}$ on all its occasions.

Variation with respect to $\lambda$ imposes the condition (\ref{definition}). Variation with respect to $\phi$ gives
\begin{equation}
\label{phieq}
\bigtriangledown_{\mu}\left(\lambda\partial^{\mu}\phi\right)=0
\end{equation}
where $\lambda\equiv\lambda^{\mu}_{\mu}\equiv g_{\mu\nu}\lambda^{\mu\nu}.$ Using (\ref{definition}), the Einstein equations read
$$G_{\mu\nu}+\lambda_{\mu\nu}=0,$$
and we see that (\ref{phieq}) is equivalent to (\ref{eom}). Finally, we vary with respect to ${\tilde g}$ and get
$$\lambda^{\mu\nu}{\tilde g}^{\alpha\beta}(\partial_{\alpha}\phi)(\partial_{\beta}\phi)-\lambda^{\rho\sigma}{\tilde g}_{\rho\sigma}{\tilde g}^{\mu\alpha}(\partial_{\alpha}\phi){\tilde g}^{\nu\beta}(\partial_{\beta}\phi)=0.$$
Together with (\ref{definition}) it gives
\begin{equation}
\label{lambdatensor}
\lambda_{\mu\nu}=\lambda(\partial_{\mu}\phi)(\partial_{\nu}\phi).
\end{equation}

Substituting the last equation into the Einstein equations and separating the trace part, we see that the effective stress tensor (\ref{effective}) from \cite{mimetic} is reproduced. For spatially homogeneous field $\phi$ we have ${\dot\phi}=0$ and the equation of motion (\ref{phieq}) implies $\lambda\propto\frac{1}{a^3}$ mimicking the Dark Matter behaviour.

Actually, we can do better about the form of the action. We see that $\lambda_{\mu\nu}$ is fully determined by its trace. Therefore, it is tempting to leave only the trace part of the constraint-fixing term in the action,
\begin{equation}
\label{final}
S=-\int\left(R(g)+\lambda\left(1-g^{\mu\nu}(\partial_{\mu}\phi)(\partial_{\nu}\phi)\right)\right)\sqrt{-g}d^4x.
\end{equation}
We claim that this is an equivalent formulation of the model at hand. Indeed, the variation with respect to $\lambda$ gives the constraint (\ref{fixednorm}) right away. The Einstein equations take the form
\begin{equation}
\label{newEinstein}
G_{\mu\nu}+\lambda(\partial_{\mu}\phi)(\partial_{\nu}\phi)=0
\end{equation}
as before. And, finally, the variation with respect to $\phi$ reproduces the equation of motion (\ref{phieq}).

\section{Some further speculations and Conclusions}

The mimetic DM proposal is equivalent to the standard GR with a scalar field which enters the action only by a constraint term demanding that the gradient is a unit time-like vector. It would be interesting to promote the Lagrange multiplier to a physical field, preferably with condition $\lambda>0$, which could act, in certain regimes, approximately in a way described above.

The most obvious way to proceed is to give the standard kinetic term $\left(\bigtriangledown\lambda\right)^2$ and a potential $V(\lambda)$ to the scalar variable $\lambda$. In a spatially homogeneous regime, the equation of motion would be ${\ddot\lambda}+3H{\dot\lambda}+V^{\prime}(\lambda)-\mu^3({\dot\phi}^2-1)=0$ where we have introduced a mass scale $\mu$ such that the Lagrange multiplier is $\mu^3\lambda$, and the field $\lambda$ has canonical dimension $+1$. If we insist to have an exact $\dot\phi=1$ solution (and non-interacting mimetic Dark Matter), then $\lambda\sim a^{-3}$ due to (\ref{phieq}), and the only way to satisfy the equation for $\lambda$ is to find a potential  for which $3\lambda{\dot H}+V^{\prime}=0$. If we assume that mimetic Dark Matter dominates over all other contributions to the stress tensor including the dynamical parts of $\lambda$, then $3H^2=\mu^3\lambda$, and we easily see that for $\lambda=\frac{C}{a^3}$ to be a solution we need $V(\lambda)=\frac{3^{5/2}\mu^{3/2}}{17C^{1/3}}\lambda^{17/6}$ which is badly contrived but not unreasonable. Moreover, one can hope for some stability of such regimes. If $\dot\phi$ becomes a bit larger than $1$, the field $\lambda$ experiences an additional effective force which tries to make it roll slower, and then the equation (\ref{phieq}), ${\dot\lambda}{\dot\phi}+3H\lambda{\dot\phi}+\lambda{\ddot\phi}=0$, produces a small negative $\ddot\phi$. We leave any detailed discussions of possible developments to other works.

It is remarkable that the conformal sector of general relativity can produce rather rich opportunities for theoretical model building if gently modified. Many years ago a unimodular gravity has been proposed to get rid off the problem of the cosmological constant. However, it was shown that the cosmological constant reappears as a constant of integration \cite{Henneaux}. Now we see that another dark sector -- pressureless dust -- can also appear via an integration constant in a model with modified variation in the conformal sector. It would be interesting to thoroughly explore possible extensions of mimetic Dark Matter model with potentially far-reaching phenomenological applications.

{\bf Acknowledgements.} The author is supported by Russian Foundation for Basic Research Grant No. 12-02-31214, and by the Saint Petersburg State University grant No. 11.38.660.2013.

\end{document}